\documentstyle{article}
\newcommand{\beq}{\begin{equation}}
\newcommand{\eeq}{\end{equation}}
\newcommand{\beqa}{\begin{eqnarray}}
\newcommand{\eeqa}{\end{eqnarray}}

\font\tenrm=cmr10
\font\tenit=cmti10
\font\elevenbf=cmbx10 scaled\magstep 1
\font\elevenrm=cmr10 scaled\magstep 1
\font\elevenit=cmti10 scaled\magstep 1

\renewenvironment{thebibliography}[1]
 { \elevenrm
   \begin{list}{\arabic{enumi}.}
    {\usecounter{enumi} \setlength{\parsep}{0pt}
     \setlength{\itemsep}{3pt} \settowidth{\labelwidth}{#1.}
     \sloppy
    }}{\end{list}}

\begin{document}
{\hfill SHEP-92/93-20\\}

{\hfill hep-ph/9305218\\}

\begin{center}
\vglue 0.6cm
{
%{\tenbf WORLD SCIENTIFIC PUBLISHING COMPANY\\}
 {\elevenbf        \vglue 10pt
               RADIATIVE CORRECTIONS TO SUPERSYMMETRIC \\
               \vglue 3pt
               HIGGS BOSON MASSES
\footnote{talk presented at SUSY-93, Boston, March 1993}
\\}
\vglue 1cm
{\elevenrm Peter L. White \\}
\baselineskip=13pt
{\tenit Physics Department, University of Southampton, \\}
\baselineskip=12pt
{\tenit Southampton S09 5NH, UK. \\}}

%\vglue 0.3cm
%{\tenrm and\\}
%\vglue 0.3cm
%{\tenrm SECOND AUTHOR'S NAME\\}
%{\tenit Group, Company, Address, City, State ZIP/Zone, Country\\}

\vglue 0.8cm
{\tenrm ABSTRACT}
\end{center}
\vglue 0.3cm
{\rightskip=3pc
 \leftskip=3pc
 \tenrm\baselineskip=12pt
 \noindent
%Typeset the abstract with an indentation
%of 3 picas on the left and right margins and in 10 point roman with
%baselineskip of 12 point.

We discuss the large radiative corrections to the masses of
supersymmetric Higgs bosons in the MSSM and in its simplest extension,
the NMSSM, with particular attention paid to the bounds on the lightest
CP-even Higgs mass found in both models. In the case of the MSSM, these
corrections are found to be primarily associated with the effects top
quark and stop squark loops, while for extended models they also include
significant contributions from Higgs and Higgsino loops. }

\vglue 0.6cm
{\elevenbf\noindent 1. Introduction - SUSY Higgs Bosons at Tree Level}
\vglue 0.2cm
{\elevenit\noindent 1.1. MSSM}
\vglue 0.1cm
\baselineskip=14pt
\elevenrm
The Minimal Supersymmetric Standard Model\cite{reviews} (MSSM) is the
simplest  possible supersymmetric theory which symmetry breaking can
reduce to the  Standard Model. It contains not only the usual fermionic
sector with SU(3)$\times$SU(2)$\times$U(1) gauge bosons plus their
superpartners, but also two supermultiplets of Higgs doublets. This is
the minimum Higgs sector which allows for the correct electro-weak
breaking structure and mass couplings, because supersymmetry both forces
the inclusion of higgsinos, which cause anomalies which cannot be
cancelled with only one doublet, and restricts the possible Yukawa
couplings between Higgses and fermions.

We shall use the convention that the two doublets are labelled $H_1$ and
$H_2$, with $H_2$ coupling to the up type quarks. The superpotential is
then given by
\beq
W = h_uQH_2u + h_dQH_1d + h_eLH_1e + \mu H_1H_2
\eeq
where we have only shown one generation. When supersymmetry is broken,
the resulting scalar potential acquires soft breaking terms of the form
\beq
V_{soft} = \sum_i m_i^2 \vert\phi_i^2\vert
     + h_uA_uQH_2u + h_dA_dQH_1d + h_eA_eLH_1e + \mu BH_1H_2
\eeq
where the sum is over all spin 0 particles, and we use the usual
convention of using the same label for the scalars as for the full
supermultiplet.

With two doublets we shall have more physical Higgs particles -- two
CP-even (labelled $h$ for the lighter and $H$ for the heavier), one
CP-odd (or axial, labelled $A$), and one charged (labelled $H^+$) -- in
additional to the usual Goldstone modes.

It is simple to derive the following Higgs potential from the lagrangian :
\beqa
V_{Higgs}     = \frac{\lambda_1}{2} (H_1^{\dag}H_1)^2
             &+& \frac{\lambda_2}{2} (H_2^{\dag}H_2)^2
              + (\lambda_3+\lambda_4) (H_1^{\dag}H_1) (H_2^{\dag}H_2)
              - \lambda_4 \vert H_2^{\dag}H_1 \vert ^2 \nonumber \\
             &+& m_1^2 (H_1^{\dag}H_1)
              + m_2^2 (H_2^{\dag}H_2)
              - m_3^2 (H_1^T i\sigma_2 H_2 + {\rm h.c.})
\eeqa
where we may use the requirements of supersymmetry to obtain
\beqa
\lambda_1=\lambda_2=\frac{g_1^2+g_2^2}{4}
\qquad
\lambda_3=\frac{g_2^2-g_1^2}{4}
\qquad
\lambda_4=\frac{g_2^2}{2}
\eeqa
and the $m_i$ are functions of the unknown soft parameters of the
theory. Thus we may calculate all of the Higgs boson masses in terms of
the (known) gauge couplings and three unknown masses. Defining
$<H_1^0>=\nu_1$, $<H_2^0>=\nu_2$, $\tan\beta=\frac{\nu_2}{\nu_1}$, and
$\nu^2=\nu_1^2+\nu_2^2$, we may use the relation
$m_z^2=\frac{1}{2}(g_1^2+g_2^2)\nu^2$ to remove one of the unknown masses.
It is then conventional to reparametrise in terms of $\tan\beta$ and
$m_A$, the mass of the CP-odd particle. Once we have done this, we find
\beqa
m_C^2&=&m_A^2+m_W^2 \nonumber \\
m_{h,H}^2 &=& \frac{1}{2}
     \Bigl( m_A^2+m_Z^2 \pm \sqrt{m_A^4+m_Z^4-
            2m_A^2m_Z^2\cos 4\beta} \Bigr)
\eeqa
where $m_C$ is the mass of the $H^+$

{}From this analysis, we should note primarily that the $H^+$ mass is
larger than the W mass; while the $h$ mass is less than both $m_A$ and
$m_Z$.  This forms an extremely firm prediction about the Higgs sector.
We should note, however, that as yet we have not discussed radiative
corrections, and are working purely with a low energy effective theory
including only Higgs bosons.

\vglue 0.2cm

{\elevenit \noindent 1.2. NMSSM}
\vglue 0.1cm

The Next-to-Minimal Supersymmetric Model\cite{NMSSM} (NMSSM) was
discussed in detail in the talk by Terry Elliott at this conference
\cite{terrytalk}. This model has a Higgs structure identical to that of
the MSSM, but with an extra singlet, and an extra term in the
superpotential  $\lambda NH_1H_2+\frac{k}{3}N^3$ to replace the $\mu$
parameter of the MSSM.

As was also described earlier at this conference
\cite{terrytalk,kanetalk}, the consideration of the perturbative
bound on the coupling $\lambda$ gives a bound on the mass of the
lightest Higgs boson which is rather larger than that found for the
MSSM\cite{NMSSMbnd}.

\vglue 0.6cm

{\elevenbf\noindent 2. Radiative Corrections}
\vglue 0.2cm

So far we have calculated masses using a low energy effective theory
from which all heavy particles (including all superpartners) have been
removed.  The first problem with this analysis is that we have used
relations defining the couplings $\lambda_i$ which are only strictly
valid at some high energy scale which we call $m_{susy}$, above which
supersymmetry is a good approximation. Below this scale, the couplings
will run from their starting values. This suggests a renormalisation
group analysis to account for the difference between the values of the
parameters at the supersymmetry and Higgs (weak) scales.

Furthermore, in decoupling the superpartners at high energy, we have
neglected any threshold effects, which may come both from the fact that
the superpartners are not all degenerate with mass $m_{susy}$ but have
some spectrum, and from the finite one-loop diagrams involving these
particles which may give non-zero contributions to the Higgs masses.
These effects can be best evaluated by either simply calculating the
diagrams at some high energy, or by carrying out a one-loop
Coleman-Weinberg effective potential calculation.

Lastly, we should also be careful that loops of light Higgs particles do
not themselves contribute substantial radiative corrections to the mass
matrices.

We shall consider only the top and stop (and, for the charged mass, the
sbottom) as coupling strongly to the Higgs sector, as all of the other
particles couple through Yukawa couplings which are relatively very
small (unless $\tan\beta >>1$ in which case the bottom quark Yukawa is
also large; we shall not consider this case). Further corrections may
also be given by loops of Higgses and Higgsinos. We shall neglect the
consideration of wave-function renormalisation and of gauge boson loops
as relatively small. The former is negligible so long as the masses
which we calculate are small relative to those of the particles in the
contributing loops.

\vglue 0.5cm
{\elevenbf \noindent 3. Results - MSSM}
\vglue 0.2cm

The situation in the MSSM is relatively simple, because as we have seen
the couplings involving Higgses and Higgsinos are small (of order
$g_2^2$ at worst), and so the most important effects are from top and
stop loops. If we begin by considering only top effects, we recall that
the top mass is given by $m_t=h_t\nu\sin\beta$, where $h_t$ is the top
quark Yukawa coupling. The simplest way of finding the top quark
contribution is to use the renormalisation group (RG) equations below
the supersymmetry scale \cite{csw}. These are rather long, but the
largest correction\cite{MSSMRG} is given by
\beq
16\pi^2\frac{\partial\lambda_2}{\partial t}= -12h_t^4 \ +\ \cdots
\eeq
where $t$ is the log of the renormalisation point. It is then trivial to
solve this reduced form to give the result
\beq
m_h^2 \leq m_Z^2 \ + \ \frac{3}{4\pi^2}h_t^4\nu_2^2\log
      \left( \frac{m_{susy}^2}{m_t^2} \right )
\eeq
which has also been obtained using effective potential techniques
\cite{erz}. In fact, solving the full RG equations shows that this
estimate is rather  conservative so that the bound should be rather
lower, because the effects proportional to $h_t^2g_2^2$ are of the
opposite sign. In any case the most important correction here being
proportional to $h_t^4$ (and thus to $m_t^4$), it becomes large very
rapidly with increasing top mass.

Typical sizes of the increase in the bound found numerically are 2, 11,
22, 35 GeV for a top mass of 120, 140, 160, 180 GeV and $m_{susy}=$1TeV,
and for the simplified analytical approximation given above are around 5
to 8 GeV more.

Next we consider the effects of squarks. These are rather messy,
and to perform this calculation we may either do an effective potential
calculation or evaluate all the appropriate graphs. The results have
been given in full by Ellis {\tenit et al}\cite{erz}, and we shall not
present them here, but merely  mention the interesting features.
Firstly, the terms which are not logarithms involving the
renormalisation scale are all proportional to $A_t+\mu\cot\beta$ to some
power. This is the off-diagonal term in the stop mass matrix (to within
factors) and so these corrections are zero unless there is mixing
between left and right-handed stops. Next, the corrections are all
proportional to $h_t^4$. Finally, as we would expect by supersymmetry,
the logarithmic terms from the divergent diagrams exactly cancel the top
corrections if the top and both stops are degenerate. After some algebra,
one can derive the following analytic formula for the bound by simply
maximising the formulae with respect to all the parameters:
\beq
m_h^2 \leq m_Z^2 \ + \ \frac{3}{4\pi^2}h_t^4\nu_2^2\log
                       \left( \frac{m_{\tilde t_2}^2}{m_t^2} \right )
          + \ \frac{9}{4\pi^2}h_t^4\nu_2^2
\label{MSSMbound}
\eeq
where we use $m_{\tilde t_2}$ to indicate the heavier of the stop mass
eigenstates. Note that the $m_{susy}$ dependence has been cancelled; for
puposes of calculating top and stop loops, $m_{susy}$ is effectively the
stop mass.

The finite terms terms are relatively insensitive to the stop masses
except indirectly through the constraint that colour must not be broken,
and so the stop mass eignvalues must be positive, which gives
$2h_t\nu_2(A_t+\mu\cot\beta)\leq(m_{\tilde t_2}^2-m_{\tilde t_1}^2)$;
however they are sensitive to $A_t$ and $A_t+\mu\cot\beta$, and so for
most arbitrarily chosen values of the parameters the lighter Higgs mass
will not approach this bound, and in fact for very large $A_t$ the
corrections to the bound may even be negative.

In summary, Eq.~(\ref{MSSMbound}) is a pessimistic view of the final
bound after radiative corrections; one would not expect it to be
realised except in the rather unlikely case that all of the parameters
are such as to make the mass as large as possible, but if this bound is
covered by the searches at LEP then we can safely rule out the MSSM.

In addition to these corrections to the bound, large corrections also
apply to the mass spectrum, particularly in the region where $\tan\beta$
is small (the bound is saturated as $\tan\beta\to\infty$;  we shall not
discuss these here in detail, except to note that they are particularly
affected by contributions from gauge loops. This is because the top
quark decouples at a scale $m_t$ (and similarly for stop squarks), while
the Higgs sector continues to contribute to the RG equations down to
lower energies.

\vglue 0.5cm

{\elevenbf \noindent 4. Results - NMSSM \hfil}
\vglue 0.2cm
Unfortunately, this rather simple analysis is drastically complicated in
the case of the NMSSM. Apart from the obvious point that $3\times 3$
matrices are not susceptible to simple diagonalisation, the spectrum is
much more complicated, and so our RG analysis approach, which implicitly
assumed that all Higgs bosons had at least approximately similar masses,
is invalidated. Apart from this, there are now large (order 1) couplings
in the Higgs and Higgsino sectors. This will mean that we must carry out
an analysis of the effects of loops of these particles.

The effects considered in the MSSM are still of course
present\cite{NMSSMRC}, and indeed the formulae for the radiative
corrections due to top loops is still valid. The stop effects for this
model have been derived elsewhere \cite{ulrich}, and their effects on
the bound have been studied \cite{ekw2}, with the result that for the
region of high top mass their effects on the bound may be substantial,
of order 20GeV; however, this region is not the one where the bound is
maximised, and at low top mass, where the bound reaches its highest
values, these corrections are negligible. The main effect is thus to
lift up  the lowest part of the $m_h-m_t$ curve, diluting the strong
$m_t$ dependence of the bound.

Now we must also consider effects of Higgs and Higgsino loops. In the
event that we make the assumption that we have a region of parameter
space in which all of the Higgs states are light (that is there are none
which differ from our renormalisation scale by more than, say, a factor
of 2 or so), then we may use our RG approach to find the radiative
corrections, and thus include all divergent diagrams. This leads to the
predicted result that the corrections proportional to $\lambda$ are in
fact quite large\cite{ekw}; the effect on the bound is to raise it by
approximately  7GeV (but by rather less for very large and small top
mass), and this result is that previously presented by Terry Elliott
\cite{terrytalk}.

In the event that the Higgs spectrum includes heavy mass eigenstates, or
if we wish to include Higgsino effects or effects from finite Higgs
loops,  the situation is much more complicated. The only technique which
we can use to deal with this case is that of Coleman-Weinberg analysis,
writing out all the tree level mass matrices and then using them to
calculate the effective potential. Unfortunately, while for scalars this
is quite simple, since the mass Higgs and Higgsino mass matrices reduce
to the usual 3$\times$3 and 2$\times$2 blocks, there is no such simple
result when we consider the charged and CP-odd particles, and we must
consider the full 10$\times$10 mass matrix. This is impossible to do
analytically, and so we must carry out the differentiations numerically.
This work is still in progress, but it seems that the radiative
corrections are typically of order a few GeV and increase the calculated
scalar masses for most values of the parameters, reaching larger values
if $r>>1$, where they may also give corrections of opposite sign.

In conclusion, it is relatively simple to find the radiative corrections
to the bound from the effects of top and stop loops, the latter being
more difficult to do with precision; but a full analysis of the Higgs
and Higgsino corrections to the bound awaits the calculation of the
analytic effective potential corrections. Furthermore, this depends on
arbitrary parameters (such as masses of Higgs bosons which are not
directly involved in the bound) to such an extent that it is difficult
to find results without making assumptions about which region of
parameter space is most interesting. However, for a given set of
parameters it is simple to find the full spectrum including radiative
corrections.

\vglue 0.5cm
{\elevenbf   \noindent 5. Conclusions \hfil}
\vglue 0.2cm

In conclusion, the effects of radiative corrections on the Higgs mass
spectrum are extremely large in both models considered here. In the MSSM
they are dominated by the well-known contributions of top and stop
loops, while for the NMSSM they also include comparably large and
(usually) positive corrections from loops of Higgses and Higgsinos;
however in the latter model it is much harder to make concrete
predictions with regard to the bound because in general these effects
are quite strongly dependent on the spectrum and thus on which region of
the very large parameter space is used for the calculations.

One question which we might ask is how dependent the results are on the
specific model considered. Clearly, going from the minimal to
next-to-minimal models has drastically increased the complexity and size
of the radiative corrections, and it seems likely that introducing
further singlets, triplets, or other structure would make things even
worse. Given the size of the corrections (and difficulty in doing
calculations because of the size of the parameter space) in the NMSSM,
we should be rather wary of bounds in such models.

\vglue 0.5cm
{\elevenbf\noindent 6. Acknowledgements \hfil}
\vglue 0.2cm
This work was done in collaboration with Terry Elliott and Steve King
at the University of Southampton.

I would like to thank Steve Kelley for a very helpful suggestion
regarding numerical effective potential techniques.

\vglue 0.5cm
{\elevenbf\noindent 7. References \hfil}
\vglue 0.2cm


\begin{thebibliography}{9}
\bibitem{reviews} H. P. Nilles, {\elevenit Phys. Rep.}
 {\elevenbf 110} (1984) 1; \\
  H. E. Haber and G. L. Kane,
    {\elevenit Phys. Rep.} {\elevenbf 117} (1985) 75; \\
  J. F. Gunion, H. E. Haber, G. L. Kane, S. Dawson,
    {\elevenit ``The Higg's Hunters Guide''}
          (Addison-Wesley, Reading MA, 1990).
\bibitem{NMSSM} J. Ellis, J.F. Gunion, H. E. Haber,
                L. Roszkowski, F. Zwirner,
   {\elevenit Phys. Rev.} {\elevenbf D39} (1989) 844.
\bibitem{terrytalk} T. Elliott, talk at this conference.
\bibitem{kanetalk} G. Kane, talk at this conference.
\bibitem{NMSSMbnd}
   J. R. Espinosa and M. Quiros,
     {\elevenit Phys. Lett.} {\elevenbf B279} (1992) 92; \\
   G. L. Kane, C. Kolda, and G. D. Wells,
     {\elevenit ``Calculable Upper Limit on the Mass of the Lightest Higgs
          Boson in Perturbatively Valid Supersymmetric Theories with
          Arbitrary Higgs Sectors''},
     Michigan preprint UM-TH-93-24,
       {\elevenit Phys. Rev. Lett.}, to be published.
\bibitem{csw} M. Carena, K. Sasaki, and C. E. M. Wagner,
 {\elevenit Nucl. Phys.} {\elevenbf B381} (1992) 66.
\bibitem{MSSMRG}
   H. Haber and R. Hempfling,
     {\elevenit Phys. Rev. Lett.} {\elevenbf 66} (1991) 1815;\\
   Y. Okada, M. Yamaguchi, and T. Yanagida,
     {\elevenit Prog. Theor. Phys.} {\elevenbf 85} (1991) 1,
     {\elevenit Phys. Lett.} {\elevenbf B262} (1991) 54;\\
   R. Barbieri, M. Frigeni, and F. Caravaglios,
     {\elevenit Phys. Lett.} {\elevenbf B258} (1991) 167; \\
   J. L. Lopez and D. V. Nanopoulos,
     {\elevenit Phys. Lett.} {\elevenbf B266} (1991) 397; \\
   A. Brignole,
     {\elevenit Phys. Lett.} {\elevenbf B277} (1992) 313,
     {\elevenit Phys. Lett.} {\elevenbf B281} (1992) 284.
\bibitem{erz} J. Ellis, G. Ridolfi, and F. Zwirner, {\elevenit Phys. Lett.}
   {\elevenbf B257} (1991) 83, {\elevenit Phys. Lett.}
   {\elevenbf B262} (1991) 477;\\
  A. Brignole, J. Ellis, G. Ridolfi, and F. Zwirner,
    {\elevenit Phys. Lett.} {\elevenbf B271} (1991) 123.\\
\bibitem{NMSSMRC}
   P. Binetruy and C. Savoy,
   {\elevenit Phys. Lett.} {\elevenbf B277} (1992) 453; \\
   U. Ellwanger and M. Rausch de Trauenberg,
   {\elevenit Z. Phys.} {\elevenbf C53} (1992) 521;\\
   U. Ellwanger and M. Lindner,
   {\elevenit ``Constraints on New Physics from the Higgs and Top Masses''},
     Heidelberg preprint HD-THEP-92-48;\\
   W. ter Veldhuis,
     {\elevenit ``Mass of the Lightest Higgs Boson in the MSSM with an
        additional Singlet''},
     Purdue preprint PURD-TH-92-11;\\
   J. R. Espinosa and M. Quiros,
     {\elevenit ``Upper bound on the Lightest Higgs Boson Mass
             in General Supersymmetric Standard Models''},
     Madrid preprint IEM-FT-64/92.
\bibitem{ulrich} U. Ellwanger,
 {\elevenit ``Radiative Corrections to the Neutral Higgs Spectrum in
                 Supersymmetry with a Singlet''},
 Heidelberg Preprint HD-THEP-93-4, {\elevenit Phys. Lett.}, to be published.
\bibitem{ekw2} T. Elliott, S. F. King, and P. L. White,
 {\elevenit ``Squark Contributions to Higgs Boson Masses in the NMSSM''},
 Southampton Preprint SHEP 92/93-18, in preparation.
\bibitem{ekw} T. Elliott, S. F. King, and P. L. White,
 {\elevenit ``Supersymmetric Higgs Bosons at the Limit''},
 Southampton Preprint SHEP 92/93-11, {\elevenit Phys. Lett.} to be published.

\end{thebibliography}
\end{document}